\documentclass[12pt]{iopart}
\usepackage{graphicx}
\begin{document}
\title[One-transit paths and steady-state of a non-equilibrium process]{One-transit paths and steady-state of a non-equilibrium process in a discrete-time update}
\author{Vahid Fayaz}
\address{Islamic Azad University, Hamedan Branch, Hamedan, Iran}
\author{Farhad H. Jafarpour}
\address{Bu-Ali Sina University, Physics Department, Hamedan, Iran}
\ead{farhad@ipm.ir}
\author{Seyedeh Raziyeh Masharian}
\address{Islamic Azad University, Hamedan Branch, Hamedan, Iran}
\author{Somayeh Zeraati}
\address{Bu-Ali Sina University, Physics Department, Hamedan, Iran}
\begin{abstract}
We have shown that the partition function of the Asymmetric Simple Exclusion Process with open boundaries in a sublattice-parallel
updating scheme is equal to that of a two-dimensional one-transit walk model defined on a diagonally rotated square lattice. It has
been also shown that the physical quantities defined in these systems are related through a similarity transformation.
\end{abstract}
\maketitle
\begin{figure}
\centering
\includegraphics[width=3in]{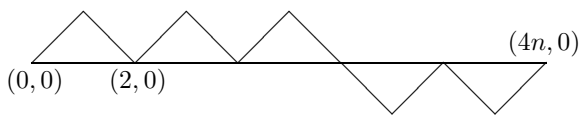}
\caption{\label{fig1} Sketch of a simple one-transit path.}
\end{figure}
\begin{figure}
\centering
\includegraphics[width=2in]{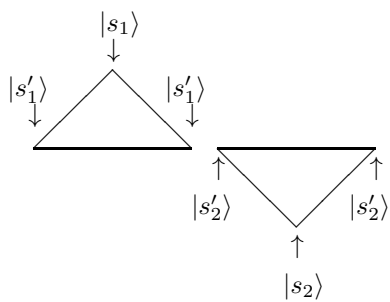}
\caption{\label{fig2} The base vectors associated with the vertices and contacts.}
\end{figure}

Recently, much attention has been devoted to the study of two-dimensional walk models and their relations with the one-dimensional driven-diffusive systems \cite{BE04,BGR04,JZ10}. The reason is that the partition function of some of these walk models can be related to those of some of the one-dimensional out-of-equilibrium driven-diffusive systems with open boundaries. One of the most simple and important system of this type is the Asymmetric Simple Exclusion Process (ASEP) with open boundaries \cite{dehp}.\\
The ASEP is a widely studied driven-diffusive system which reveals a wealth of interesting critical phenomena. The ASEP is defined on a one-dimensional discrete lattice. The classical particles are allowed to enter the lattice only from the left boundary and leave the lattice only from the right boundary. The particles also hop unidirectionally toward the right boundary. It is known that a single product shock measure has a simple random walk dynamics in this system, without any constraints on the parameters of the system, provided that we consider a discrete-time updating scheme. In this case the particles are injected into the leftmost lattice site with the probability $\alpha$ if the target site is empty. They are also extracted from the rightmost lattice site with the probability $\beta$ if it is already occupied. We assume that in the bulk of the lattice a particle at a given site $k$ deterministically move to the site $k+1$ provided that it is not occupied by other particles, which means that double occupation is prohibited.\\
In this paper we consider the ASEP with open boundaries in the sublattice-parallel update \cite{SCH93,HH96}. We are looking for a two-dimensional walk model which shares similar critical properties with the ASEP. It turns out that there exists such a two-dimensional walk model as we will describe it later. These systems have equal partition functions and the physical quantities in these systems are related through a simple transformation.\\
The ASEP is defined on a lattice of length $2n$. The time evolution of the probability distribution vector $\vert P(t) \rangle$ is given by the following master equation:
\begin{equation}
\label{time evol}
T \vert P(t) \rangle
= \vert P(t+1) \rangle.
\end{equation}
The sublattice-parallel updating scheme is defined as follows: We divide the bulk dynamics consists into two half time steps. In the first half time step the pairs of neighboring sites ($2k,2k+1$) for $k=1,\cdots,n-1$ and also the first and the last lattice sites are updated. In the second half time step the pairs of neighboring sites ($2k-1,2k$) for $k=1,\cdots,n$ are updated. Considering this definition, the transfer matrix $T$ in (\ref{time evol}) is given by the multiplication of two factors $T=T_2 T_1$ defined as:
\begin{eqnarray}
\label{T1}
T_1 &=& {\cal L} \otimes {\cal T} \otimes \ldots \otimes
        {\cal T} \otimes {\cal R} \,\;=\;\,
{\cal L} \otimes {\cal T}^{\otimes (n-1)} \otimes {\cal R} \nonumber\\
\label{T2}
T_2 &=& \hspace{3.8mm} {\cal T} \otimes {\cal T} \otimes \ldots \otimes
        {\cal T} \hspace{3.8mm} \,\;=\;\,  {\cal T}^{\otimes n} \nonumber
\end{eqnarray}
where ${\cal T}$, ${\cal L}$ and ${\cal R}$ are given by:
\begin{equation}
\label{trans mat}
{\cal T} \;=\; \left(
\begin{array}{cccc}
1 & 0 & 0 & 0 \\
0 & 1 & 1 & 0 \\
0 & 0 & 0 & 0 \\
0 & 0 & 0 & 1
\end{array} \right),\;
{\cal L} \;=\; \left(
\begin{array}{cc}
1-\alpha & 0 \\
\alpha & 1
\end{array} \right),\;
{\cal R} \;=\; \left(
\begin{array}{cc}
1 & \beta \\
0 & 1-\beta
\end{array} \right).
\end{equation}
The matrix $\cal T$ is written in the basis $(00,01,10,11)$ when $0$ stands for an empty lattice site and $1$ stands for an occupied lattice site. The matrices $\cal L$ and $\cal R$ are also written in the basis $(0,1)$. \\
Following \cite{JM09} we define two different types of product shock measures $\vert \mu_{2k}\rangle$ ($k=1,\cdots,n$) and  $\vert \mu_{2k+1} \rangle$ ($k=0,\cdots,n$) as follows:
\begin{equation}
\label{shocks}
\fl
\begin{array}{l}
\vert \mu_{2k} \rangle=
\left(\begin{array}{c}
1-\hat{\rho}_1 \\ \hat{\rho}_1
\end{array}\right)\otimes
\left(\begin{array}{c}
1-\rho_{1} \\ \rho_{1}
\end{array}\right)\otimes \cdots \otimes
\underbrace{ \left(\begin{array}{c}
1-\rho_{2} \\ \rho_{2}
\end{array}\right)}_{2k}\otimes
\left(\begin{array}{c}
1-\hat{\rho}_2 \\ \hat{\rho}_2
\end{array}\right)\otimes \cdots \otimes
\underbrace{\left(\begin{array}{c}
1-\rho_{2} \\ \rho_{2}
\end{array}\right)}_{2n},
\\
\vert \mu_{2k+1} \rangle=
\left(\begin{array}{c}
1-\hat{\rho}_1 \\ \hat{\rho}_1
\end{array}\right)\otimes
\left(\begin{array}{c}
1-\rho_{1} \\ \rho_{1}
\end{array}\right)\otimes \cdots \otimes
\underbrace{ \left(\begin{array}{c}
1-\hat{\rho}_2 \\ \hat{\rho}_2
\end{array}\right)}_{2k+1}\otimes
\left(\begin{array}{c}
1-\rho_2 \\ \rho_2
\end{array}\right)\otimes\cdots \otimes
\underbrace{\left(\begin{array}{c}
1-\rho_{2} \\ \rho_{2}
\end{array}\right)}_{2n}.
\end{array}
\end{equation}
Note that the position of the shock in $\vert \mu_{2k}\rangle$ lies between two consecutive sites $2k-1$ and $2k$. At the same time the shock position lies between two consecutive sites $2k$ and $2k+1$ in $\vert \mu_{2k+1}\rangle$. The authors in \cite{JM09} have shown that the time evolution of the shock position in (\ref{shocks}) governed by (\ref{time evol}) is similar to that of a simple random walker. The shock position hops to the left and to the right with the probabilities $\delta_{l}$ and $\delta_{r}$ respectively provided that:
\begin{equation}
\label{Densities}
\begin{array}{l}
\hat{\rho}_1=0 \; , \; \hat{\rho}_2=1-\beta\\
\rho_1=\alpha \; , \; \rho_2=1\\
\delta_{r}=\beta \; , \; \delta_{l}=\alpha.
\end{array}
\end{equation}
The steady-state of the system $\vert P^{\ast}\rangle$ which obeys $\vert P^{\ast}\rangle=T\vert P^{\ast}\rangle$ can be written as a superposition of the shocks defined in (\ref{shocks}) as follows:
\begin{equation}
\label{SS1}
\vert P^{\ast} \rangle = \frac{1}{Z}\sum_{k=1}^{2n+1}c_k \vert \mu_k \rangle
\end{equation}
provided that:
\begin{eqnarray}
&& c_{2k}=\delta_{s}\Big(\frac{\delta_{l}}{\delta_{r}}\Big)^{2(n-k)+1} \; \mbox{for} \; k=1,\cdots,n\\
&& c_{2k+1}=\Big(\frac{\delta_{l}}{\delta_{r}}\Big)^{2(n-k)} \; \mbox{for} \; k=0,\cdots,n
\end{eqnarray}
in which $\delta_{s}=1-\delta_{r}-\delta_{l}$. The partition function of the system $Z$ (which is also the normalization factor) is also given by:
\begin{equation}
\label{Z1}
Z=\sum_{k=1}^{2n+1}c_k=\frac{\delta_{r}(1-\delta_{l})}{\delta_{r}-\delta_{l}}-
\frac{\delta_{l}(1-\delta_{r})}{\delta_{r}-\delta_{l}}\Big(\frac{\delta_{l}}{\delta_{r}}\Big)^{2n}.
\end{equation}
In the steady-state and depending on whether $\delta_{l}$ is larger or smaller than $\delta_{r}$, the system undergoes a phase transition. The coexistence line
occurs where $\delta_{l}=\delta_{r}$. In both phases the density of particles on the lattice have exponential behaviors while on the coexistence line it is linear because of considering a superposition of shocks with unbiased random walk dynamics.\\
In \cite{HH96} the author has rigorously shown that the steady-state of the ASEP in a sublattice-parallel update can be obtained using a matrix product method recently reviewed in \cite{BE07}. According to this method we assign an operator to each state of a lattice site in the steady-state. The two operators $D$ and $E$ are associated with the presence of a particle and a vacancy at an even lattice site respectively. The two operators $\hat{D}$ and $\hat{E}$ are also associated with the presence of a single particle and a vacancy at an odd lattice site respectively. For a system of length $2n$ we associate a product of $2n$ operators to each configuration of the system consisting of particles and vacancies at different lattice sites. The unnormalized weight associated with any configuration can be obtained by considering a matrix element of this product. \\
Using the matrix product method and in the same basis mentioned above one can rewrite the steady-state of the ASEP as follows:
\begin{equation}
\label{SS2}
\vert P^{\ast}\rangle= \frac{1}{Z} \langle \langle W \vert
\left[\left( \begin{array}{c}
{\hat E} \\ {\hat D} \end{array} \right) \otimes \left( \begin{array}{c}
E \\ D \end{array} \right)\right]^{\otimes n}
\vert V \rangle \rangle
\end{equation}
in which the normalization factor is given by:
\begin{equation}
\label{Z2}
Z=\langle \langle W \vert ({\hat E}+{\hat D})^n(E+D)^n \vert V\rangle \rangle.
\end{equation}
It can be verified that the following two-dimensional matrix representation generates the same steady-state introduced in (\ref{SS1}):
\begin{equation}
\begin{array}{l}
\label{MatrixRep}
\hat{D} = \left( \begin{array}{cccc}
\hat{\rho}_{1}& & & -\hat{d}_{0} \\
0 & & & \frac{\delta_{l}}{\delta_r}\hat{\rho}_{2}
\end{array} \right) \; , \;
\hat{E} = \left( \begin{array}{cccc}
1-\hat{\rho}_{1}& & & \hat{d}_{0} \\
0 & & & \frac{\delta_{l}}{\delta_{r}}(1-\hat{\rho}_{2})
\end{array} \right) \; ,\; \\ \\
D = \left( \begin{array}{cccc}
\rho_{1}  & & & -d_{0} \\
0 & & & \frac{\delta_{l}}{\delta_{r}}\rho_{2}
\end{array} \right) \; , \;
E = \left( \begin{array}{cccc}
1-\rho_{1} & & & d_{0}  \\
0 & & & \frac{\delta_{l}}{\delta_{r}}(1-\rho_{2})
\end{array} \right) \; , \; \\ \\
\langle\langle W \vert = (w_1, w_2) \; , \;
\vert V \rangle \rangle= \left( \begin{array}{c}
v_1 \\ v_2 \end{array} \right)
\end{array}
\end{equation}
given that:
\begin{equation}
\label{Conditions1}
\frac{v_{2}}{v_{1}} d_{0}=-\frac{(1-\delta_{r})(\rho_{2}-\rho_{1})}{(1-\delta_{l})}\Big( \frac{\delta_{l}}{\delta_{r}} \Big) \;,\;
\frac{w_{1}}{w_{2}} \hat{d}_{0}=\frac{(1-\delta_{l})(\hat{\rho}_{2}-\hat{\rho}_{1})}{(1-\delta_{r})}.
\end{equation}
Note that the two-dimensional matrix representation (\ref{MatrixRep}) also generates the same partition function (\ref{Z1}) using (\ref{Z2}). In comparison to the matrix representation introduced in \cite{JM09} we have adopted an upper triangular matrix representation.\\
In what follows we will define a one-transit walk model consisting of $4n$ steps on a diagonally rotated square lattice and explain how the partition function of this system can be related to that of the ASEP studied above. We will then discuss how the physical quantities in both systems are related through a similarity transformation.\\
Consider a random walker moving on a two-dimensional diagonally rotated square lattice. The random walker starts from the origin $(0,0)$ and takes $4n$ consecutive steps until it gets to the point $(4n,0)$. The random walker moves in the north-east (NE) or the south-east (SE) direction. The random walker never takes two consecutive steps in the NE direction along the path; however, it can take two consecutive steps in the SE direction only once during the journey. In this case the random walker moves along a path which crosses the horizontal axis only once. A typical one-transit path is given in Figure (\ref{fig1}).
Let us assume that the random walker moves along a weighted path. We assign a fugacity to the steps that the random walker takes as follows:
\begin{itemize}
\item The fugacity $\hat{z}_{1}$ to each step in SE direction and the fugacity $\hat{z}_{2}$ to each step in NE direction if the ending point is $(2k,0)$ for an odd $k$ ($k=1,3,\cdots,2n-1$).
\item The fugacity $z_{1}$ to each step in SE direction and the fugacity $z_{2}$ to each step in NE direction if the ending point is $(2k,0)$ for an even $k$ ($k=2,4,\cdots,2n$).
\item The fugacity $1$ to all other steps.
\end{itemize}
In order to find the weight of a given path one can simply multiply the fugacities of different steps of the path. The partition function of this system can be easily calculated and one finds:
\begin{equation}
\label{Z3}
Z(z_{1},\hat{z}_{1},z_{2},\hat{z}_{2})=(z_{1}\hat{z}_{1}z_{2}\hat{z}_{2})^{n}\tilde{Z}(z_{1},\hat{z}_{1},z_{2},\hat{z}_{2})
\end{equation}
in which:
\begin{equation}
\label{Z4}
\tilde{Z}(z_{1},\hat{z}_{1},z_{2},\hat{z}_{2})=\Big( \frac{\frac{1}{z_{1}\hat{z}_{1}}+\frac{1}{z_{1}\hat{z}_{2}}}{\frac{1}{z_{1}\hat{z}_{1}}-\frac{1}{z_{2}\hat{z}_{2}}} \Big) (z_{1}\hat{z}_{1})^{-n}-\Big(  \frac{\frac{1}{z_{2}\hat{z}_{2}}+\frac{1}{z_{1}\hat{z}_{2}}}{\frac{1}{z_{1}\hat{z}_{1}}-\frac{1}{z_{2}\hat{z}_{2}}} \Big) (z_{2}\hat{z}_{2})^{-n}.
\end{equation}
The first term in (\ref{Z3}) does not play any important role in the critical behavior of the system. The second term in (\ref{Z3}), which is given explicitly in (\ref{Z4}), can be reinterpreted as the partition function of a two-dimensional walk model on a diagonally rotated square lattice; however, one should assign the fugacities to the contact points (instead of the steps) in a different way. We assign a fugacity $1/\hat{z}_{1}$ to each contact point $(2k,0)$ for an even $k$ and a fugacity $1/z_{1}$ to each contact point $(2k,0)$ for an odd $k$ if the contact point is above the horizontal axis. We also assign a fugacity $1/\hat{z}_{2}$ to each contact point $(2k,0)$ for an even $k$ and a fugacity $1/z_{2}$ to each contact point $(2k,0)$ for an odd $k$ if the contact point is below the horizontal axis. In all cases we do not assign any fugacity to the contact points with the horizontal axis if they come from the first upward step or the first downward step. The partition function of this walk model given in (\ref{Z4}) can now be written using a transfer matrix method as follows:
\begin{equation}
\label{TMZ}
\tilde{Z}(z_{1},\hat{z}_{1},z_{2},\hat{z}_{2})=\langle L \vert T^{n} \vert R \rangle
\end{equation}
in which the transfer matrix $T$ is a two-step transfer matrix defined as $T=T_{o}T_{e}$. Note that $T_{o}$ and $T_{e}$ are each related to two consecutive steps i.e. $T_{o}$ ($T_{e}$) is the transfer matrix of two consecutive steps ending to the point $(2k,0)$ for an odd (even) $k$; therefore, one can rewrite $T_{o}$ and $T_{e}$ as:
\begin{equation}
\label{TMS}
T_{o}=T_{o}^{o}T_{o}^{e} \; , \; T_{e}=T_{e}^{o}T_{e}^{e}.
\end{equation}
By assigning four base vectors to the vertices and contact points of a path one can find a matrix representation for the transfer matrices and the vectors introduced in (\ref{TMZ}) and (\ref{TMS}). The vectors $\vert s_{k} \rangle$ and $\vert s'_{k} \rangle$ for $k=1,2$ are assigned to the vertices and the contact points of the path as shown in Figure (\ref{fig2}).
In an appropriate basis one can write:
\begin{eqnarray}
\vert s_{1} \rangle=\vert s'_{1} \rangle=
\left( \begin{array}{c}
1\\
0
\end{array} \right)
\; , \; \vert s_{2} \rangle=\vert s'_{2} \rangle=
\left( \begin{array}{c}
0\\
1
\end{array} \right)
\end{eqnarray}
with the following completeness property:
\begin{equation}
\sum_{k=1,2}\vert s_{k} \rangle \langle s_{k} \vert=\sum_{k=1,2}\vert s'_{k} \rangle \langle s'_{k} \vert=\cal{I}
\end{equation}
in which $\cal I$ is a $2\times 2$ identity matrix. Considering the definition of the weighted walk model the only nonzero matrix elements are:
\begin{equation}
\label{Trans}
\begin{array}{llll}
\langle s_{1} \vert T_{o}^{e} \vert s'_{1} \rangle=\frac{1}{z_1}& , & \langle s_{1} \vert T_{e}^{e} \vert s'_{1} \rangle=\frac{1}{\hat{z}_1} & , \\
\langle s'_{1} \vert T_{o}^{o} \vert s_{1} \rangle=1            & , & \langle s'_{1} \vert T_{e}^{o} \vert s_{1} \rangle=1 & , \\
\langle s'_{1} \vert T_{o}^{o} \vert s_{2} \rangle=1            & , & \langle s'_{1} \vert T_{e}^{o} \vert s_{2} \rangle=1 & , \\
\langle s_{2} \vert T_{o}^{e} \vert s'_{2} \rangle=\frac{1}{z_2}& , & \langle s_{2} \vert T_{e}^{e} \vert s'_{2} \rangle=\frac{1}{\hat{z}_2} & , \\
\langle s'_{2} \vert T_{o}^{o} \vert s_{2} \rangle=1            & , & \langle s'_{2} \vert T_{e}^{o} \vert s_{2} \rangle=1 & , \\
\langle s'_{1} \vert R \rangle=1                                & , & \langle s'_{2} \vert R \rangle=1  & , \\
\langle L \vert s'_{1} \rangle=1                                & , & \langle L       \vert s'_{2} \rangle=0  & .
\end{array}
\end{equation}
All other matrix elements are zero. Now using (\ref{Trans}) one can easily construct the transfer matrices in the above mentioned basis:
\begin{equation}
\label{Rep1}
T_{o}^{o} = T_{e}^{o}= \left( \begin{array}{cc}
1  & 1 \\
0 & 1
\end{array} \right), \;
T_{o}^{e} =\left( \begin{array}{cc}
\frac{1}{z_{1}}  & 0 \\
0 & \frac{1}{z_{2}}
\end{array} \right), \;
T_{e}^{e} =\left( \begin{array}{cc}
\frac{1}{\hat{z}_{1}}  & 0 \\
0 & \frac{1}{\hat{z}_{2}}
\end{array} \right)
\end{equation}
and the vectors $\langle L \vert$ and $\vert R \rangle$ are:
\begin{equation}
\label{Rep2}
\langle L \vert = \left( \begin{array}{cc}
1  & 0
\end{array} \right), \;
\vert R \rangle = \left( \begin{array}{c}
1  \\
1
\end{array} \right).
\end{equation}
Before going any farther let us investigate the asymptotic behavior of the partition function of the one-transit walk model given in (\ref{Z4}) in the large-$n$ limit.  It can be easily seen that in the large-$n$ limit the partition function shows two different behaviors depending on whether $z_{1}\hat{z}_{1}>z_{2}\hat{z}_{2}$ or $z_{1}\hat{z}_{1}<z_{2}\hat{z}_{2}$ which results in two different phases. This means that the phase diagram of the walk model depends only on $z_{1}\hat{z}_{1}$ and $z_{2}\hat{z}_{2}$ and not the four independent fugacities. It turns out that for the case $z_{1}\hat{z}_{1}>z_{2}\hat{z}_{2}$ most of the contacts with the horizontal axis will be from below while in the case $z_{1}\hat{z}_{1}<z_{2}\hat{z}_{2}$ they will be from above. On the coexistence line where $z_{1}\hat{z}_{1}=z_{2}\hat{z}_{2}$ the number of contacts both from above and below the horizontal axis
will vary linearly along the horizontal axis.\\
At this point we show that how the partition function of the walk model (\ref{Z4}) is related to that of the ASEP given in (\ref{Z1}). It will also leads us to understand how the physical quantities in these systems are related. Let us assume that:
\begin{equation}
z_{1}\hat{z}_{1}=1,\; z_{2}\hat{z}_{2}=(\frac{\delta_{r}}{\delta_{l}})^{2},\; z_{1}\hat{z}_{2}=\frac{(\frac{\delta_{r}}{\delta_{l}})}{1-\delta_{r}-\delta_{l}}.
\end{equation}
Now the following similarity transformation connects the two above mentioned systems:
\begin{equation}
\label{ST}
\langle\langle W \vert U=\langle L \vert,\;\;U^{-1} C^2 U=T,\;\;U^{-1} \vert V \rangle \rangle = \vert R \rangle
\end{equation}
in which:
\begin{eqnarray}
\label{U}
U = \left( \begin{array}{cc}
1  &  \frac{1-\delta_{r}}{1-\Big(\frac{\delta_{l}}{\delta_{r}}\Big)}\Big( \frac{\delta_{l}}{\delta_{r}}\Big) \\
0 &   u
\end{array} \right).
\end{eqnarray}
It is also necessary to have the following relations:
\begin{equation}
\label{Conditions2}
w_{1}=1 \;,\; w_{2}=-\frac{1-\delta_{r}}{1-\Big(\frac{\delta_{l}}{\delta_{r}}\Big)}\Big( \frac{\delta_{l}}{\delta_{r}}\Big)u^{-1} \;,\;v_{1}=\frac{1-\delta_{l}}{1-\Big(\frac{\delta_{l}}{\delta_{r}}\Big)} \;,\;
v_{2}=u.
\end{equation}
provided that $v_{1},\, v_{2},\, w_{1}$ and $w_{2}$ satisfy (\ref{Conditions1}) which is achievable since $u$, $d_{0}$ and $\hat{d}_{0}$ are free to be chosen. It is now easy to check that the partition functions of the two systems in different descriptions given by (\ref{Z1}), (\ref{Z2}) and (\ref{Z4}) are equal.\\
Let us now investigate how the physical quantities in the walk model are related to those in the ASEP. We start with the probability of finding a contact at
the lattice site $(2k,0)$ for an even $k$ from above $P_{2k}^{above}$ or below $P_{2k}^{below}$ the horizontal axis which is given by the following expressions:
\begin{eqnarray}
\label{Contacts1}
P_{2k}^{above}=\frac{\langle L \vert T^{\frac{k}{2}} \hat{A}_{2k} T^{n-\frac{k}{2}} \vert R \rangle}{\langle L \vert T^{n} \vert R \rangle}\;\; k=2,4,6,\cdots,2n,\\
\label{Contacts2}
P_{2k}^{below}=\frac{\langle L \vert T^{\frac{k}{2}} \hat{B}_{2k} T^{n-\frac{k}{2}} \vert R \rangle}{\langle L \vert T^{n} \vert R \rangle}\;\;
k=2,4,6,\cdots,2n
\end{eqnarray}
in which $\hat{A}_{2k}=\vert s_{1} \rangle_{2k-1}\;{}_{2k}\langle s'_{1} \vert$ and $\hat{B}_{2k}=\vert s_{2} \rangle_{2k-1}\;{}_{2k}\langle s'_{2} \vert$ are the contact operators at the lattice site $(2k,0)$ from above and below the horizontal axis, respectively. Using the matrix representations (\ref{Rep1}) and (\ref{Rep2}), it is easy to calculate the expressions (\ref{Contacts1}) and (\ref{Contacts2}) analytically.\\
One can also calculate the probability of finding a contact at the lattice site $(2k,0)$ for an odd $k$ from above $\tilde{P}_{2k}^{above}$ or below $\tilde{P}_{2k}^{below}$ the horizontal axis which is given by the following expressions:
\begin{eqnarray}
\label{Contacts3}
\tilde{P}_{2k}^{above}=\frac{\langle L \vert T^{\frac{k-1}{2}} T_{o} \hat{A}_{2k} T_{e} T^{n-\frac{k+1}{2}} \vert R \rangle}{\langle L \vert T^{n} \vert R \rangle}\;\;k=1,3,5,\cdots,2n-1,\\
\label{Contacts4}
\tilde{P}_{2k}^{below}=\frac{\langle L \vert T^{\frac{k-1}{2}} T_{o} \hat{B}_{2k} T_{e} T^{n-\frac{k+1}{2}}  \vert R \rangle}{\langle L \vert T^{n} \vert R \rangle}\;\;k=1,3,5,\cdots,2n-1.
\end{eqnarray}
The contact operators $\hat{A}_{2k}$ and $\hat{B}_{2k}$ have the same definitions mentioned above.\\
Now we show that how the density of the particles in the ASEP is related to the probabilities of contacts given in (\ref{Contacts1})-(\ref{Contacts4}). According to the matrix product formalism the density of the particles in an even site $\langle \tau_{k} \rangle$ ($k$ even) in the ASEP in the sublattice-parallel update is given by:
\begin{equation}
\label{Evenden}
\langle \tau_{k} \rangle=\frac{\langle\langle W \vert C^{k-1}DC^{2n-k}\vert V \rangle\rangle}{\langle\langle W \vert C^{2n}\vert V \rangle\rangle} \; \; \mbox{for} \; \; k=2,4,6,\cdots,2n.
\end{equation}
Using (\ref{ST}), (\ref{Contacts1}) and (\ref{Contacts2}) and after some straightforward calculations it turns out that (\ref{Evenden}) can be written as:
\begin{equation}
\label{deneven}
\langle \tau_{k} \rangle=\rho_{1}P_{2k}^{above}+\rho_{2}P_{2k}^{below} \; \; \mbox{for} \; \; k=2,4,6,\cdots,2n.
\end{equation}
On the other hand, in the same system and using the matrix product formalism the density of the particles in an odd site $\langle \tau_{k} \rangle$ ($k$ odd) is given by:
\begin{equation}
\label{Oddden}
\langle \tau_{k} \rangle=\frac{\langle\langle W \vert C^{k-1}\hat{D}C^{2n-k}\vert V \rangle\rangle}{\langle\langle W \vert C^{2n}\vert V \rangle\rangle} \; \; \mbox{for} \; \; k=1,3,5,\cdots,2n-1.
\end{equation}
Using (\ref{ST}), (\ref{Contacts3}) and (\ref{Contacts4}) it is not difficult to show that (\ref{Oddden}) can be written as:
\begin{equation}
\label{denodd}
\langle \tau_{k} \rangle=\hat{\rho}_{1}\tilde{P}_{2k}^{above}+\hat{\rho}_{2}\tilde{P}_{2k}^{below} \; \; \mbox{for} \; \; k=1,3,5,\cdots,2n-1.
\end{equation}
In \cite{SCH93,HH96} the authors have already studied both the phase diagram and also the behaviors of (\ref{deneven}) and (\ref{denodd}) in the large-$n$ limit; therefore, we do not repeat their results here and the reader can directly refer to those references. On the other hand, a description of the phase diagram of the same system based on the traveling shocks has been done in \cite{JM09}.\\
In this paper we aimed to show that the steady-state of the ASEP with open boundaries in a sublattice-parallel update can be described using a two-dimensional walk model. We showed that the partition functions of these models are equal by properly defining the fugacities assigned to the steps (or the contact points) of the random walker. On the other hand, the probability of finding a particles at a given lattice site in the ASEP is connected to the probability that the Dyck path touches the horizontal axis at that point. This is valid for both even and odd lattice sites as obtained in (\ref{deneven}) and (\ref{denodd}).


\end{document}